\documentclass[prd,twocolumn,aps,epsbox,groupedaddress,eqsecnum]{revtex4}
\usepackage[dvips]{graphicx}
\usepackage{bm}
\usepackage{color}
\usepackage{epsfig}
\usepackage{here}
\usepackage{subfigure}
\usepackage{amsmath}
\begin{document}
\title{Fluctuations of the flux of energy on the apparent horizon}
\author{Jos\'e P.~Mimoso}
\email{jpmimoso@fc.ul.pt}
 \affiliation{Faculdade de Ci\^{e}ncias, Departamento de F\'{i}sica and Instituto
 de Astrof\'{\i}sica  e Ci\^{e}ncias do Espa\c{c}o\\
 Universidade do Lisboa, Ed. C8, Campo Grande, 1749-016 Lisboa, Portugal.}
\author{Diego Pav\'on}
\email{diego.pavon@uab.es} \affiliation{Departamento de
F\'{\i}sica, Universidad Aut\'onoma de  Barcelona, Facultad de
Ci\'{e}ncias,\\ 08193 Bellaterra, Barcelona, Spain.}

\begin{abstract}
 Adopting the Landau-Lifshiftz method of classical fluctuations
we determine the statistical average strength of the fluctuations
of the energy flux on the apparent horizon of a homogeneous and
isotropic universe described by Einstein gravity. We find that the
fluctuations increase with the temperature of the horizon and
decrease with its area, in accordance with the features of systems
where gravity can be neglected. We further find, on the one hand,
that the  fluctuations vanish in the cosmological constant
dominated de Sitter expansion and, on the other hand, that the
domination of phantom fields is excluded. The reasonableness of
the results we  have obtained lend support to the view that the
Universe behaves as a normal thermodynamic system.
\end{abstract}

\maketitle
\section{Introduction}
Nowadays, the existence of a close connection between gravity and
thermodynamics is widely acknowledged \textemdash see for instance
\cite{Padmanabhan:2009vy} and references therein. It may be said
that this interplay was first intimated by Tolman's law for the
equilibrium temperature in a medium placed in a gravitational
field \cite{tolman} and shortly afterward by the realization that
a heat flux must run through an accelerated body in direction
opposite to the acceleration \cite{eckart}. Both effects are a
direct consequence of the equivalence principle \cite{pla-diego}.
The understanding that the said  connection is deep, and not
merely coincidental, was strongly reinforced by the discovery that
black holes obey the thermodynamic laws \cite{jakob, steve,
werner,Hayward:1997jp} and, later on, practically confirmed by the
finding that Einstein field equations can be derived from the
definition of entropy and the proportionality between the latter
and the horizon area \cite{ted}.
\\  \

In view of the above and, on the other hand, given the huge number
of degrees of freedom of the Universe one may wonder whether the
latter can be considered a thermodynamic system. Recently, this
was partially answered in the affirmative by the suggestion, based
on the observed evolution of the Hubble factor, that the entropy
of the Universe tends to a finite maximum \cite{grg-dp,
prd-mimoso} \textemdash like any other macroscopic isolated
physical system. However, it is not at all simple to determine
experimentally the evolution of the said factor \textemdash
\cite{apj-farook} and references therein. Therefore, if one wishes
to answer this question before more abundant data and data of much
higher quality become available, it seems advisable to resort to
the study the thermodynamic fluctuations of the energy flux.
\\  \

As  is well known, physical quantities of macroscopic systems
experience small random fluctuations around their average values,
because of the discontinuous nature of matter and of the thermal
motion of its microscopic constituents. They are spontaneous,
ubiquitous, grow in size with temperature and are at the root of
the inevitable, but usually controllable, ``noise" in measurement
devices. Paradigmatic examples of thermal fluctuations are, for
instance, the Brownian motion of a solid small particle in a fluid
\cite{fls, reif} and the fluctuations of the electric voltage in a
resistor \cite{reif}.
\\  \

In this work, we consider a homogeneous and isotropic universe
described at large scale by the Friedmann-Robertson-Walker (FRW)
metric, and we calculate the average size of the fluctuations of
the energy flux on the apparent horizon using the method of Landau
and Lifshitz \cite{ll,1959ll,1969ll,plb653}, succinctly recalled
below. In principle, one might study the said fluctuations on any
other closed surface as the event horizon. However, the former
horizon is decidedly more suitable as it fulfills the laws of
thermodynamics, while the latter does not \cite{bga}.
\\   \

We resort to the method of Landau and Lifshitz (LL) to determine
the strength of the fluctuations which, although being based on
microscopic considerations, offers a macroscopic approach to the
issue and thus has the great advantage that it obviates the use of
concepts  stemming from a microscopic description, e.g.,
distribution functions, when the latter are unclear. As  we aim at
assessing the fluctuation of the energy flux across the apparent
horizon of the expanding Universe, the LL method  does not depend
on any underlying microscopic entities making up the spacetime and
thus avoids such an unclear issue. As it is, the problem only
involves  tackling the flux of matter, radiation and/or dark
energy,  which are  the familiar components that source the
gravitational field, and hence the usual Einstein equations are
valid in this classical context.
\\  \

We wish to emphasize that the behavior of the fluctuations of a
physical system gives information about the properties of the
latter. If the fluctuations of the energy flux mentioned above do
behave in accordance with the fluctuations in normal systems which
are not dominated by gravity, our confidence in the Universe being
indeed a thermodynamic system (one that complies with the
thermodynamic laws) will get significantly strengthened.
\section{Fluctuations on the apparent horizon}
At this point, it is expedient to recall the notion of apparent
horizon in a FRW universe \textemdash see Refs. \cite{cqg-bak-rey}
and \cite{cai-cao} for details (or see also Refs.
\cite{Mimoso:2016jwg,Faraoni:2015ula,Binetruy:2014ela} for
additional insights). A spherically symmetric spacetime region
will be called  ``trapped"  if the expansion of ingoing and
outgoing null geodesics,  normal to the spatial two-sphere of
radius $\tilde{r}$ [where $\tilde{r} = a(t) \, r$] centered at the
origin (i.e., at the comoving observer), is negative. By contrast,
the region will be called ``antitrapped" if the expansion of the
geodesics is positive. In normal regions outgoing null rays have
positive expansion and ingoing null rays, negative expansion.
Thus, the antitrapped region is given by the condition
 \begin{equation}
 \tilde{r} > \frac{1}{\sqrt{H^2+\frac{k}{a^2} }}\; ,
 \end{equation}
where $\, H$ and $ \, k$ stand for the Hubble rate and the spatial
curvature index. Clearly, the surface of the apparent horizon is
nothing but the boundary hypersurface of the  spacetime
antitrapped region. In the case of an exact de Sitter expansion,
the apparent and event horizons coincide.
\\  \

Since the radius of the apparent horizon fulfills $\, \tilde{r}_H
= 1/\sqrt{H^{2} \, + \, k a^{-2}}$, the area and entropy of the
horizon, in units of the Boltzmann's constant, are
\cite{cqg-bak-rey, cai-cao}
\begin{equation}
{\cal A}_{H} = 4 \pi \tilde{r}^{2} _{H} = \frac{4 \pi}{H^{2}\, +
\,\frac{k}{a^{2}}}\, \quad {\rm and} \quad S_{H} =
\frac{1}{\ell^{2}_{P}} \frac{ \pi}{H^{2}\, + \,\frac{k}{a^{2}}}\,
, \label{eq:area-entropy}
\end{equation}
respectively.
\\  \

\noindent As the Universe expands at the Hubble rate the energy
inside the horizon increases at a rate
\begin{equation}
- \dot{E} = {\cal A}_{H} (\rho \, + \, p) H \, \tilde{r}_{H} = -
\frac{{\cal A}_{H}}{4 \pi G} \, \left(\dot{H} \, - \,
\frac{k}{a^{2}} \right)\, \frac{H}{\sqrt{H^{2} \, + \, k
a^{-2}}}.
 \label{eq:dotE}
\end{equation}
In arriving at the second equality, we relied on the conservation
of matter energy,
\begin{equation}
\dot\rho+3H\,(\rho+p)=0\; ,
\end{equation}
alongside Friedmann's equation
\begin{equation}
3 \left(H^{2}\, + \, \frac{k}{a^{2}}\right)=8\pi G\,\rho .
\label{eq:friedmann}
\end{equation}
\\  \

In (\ref{eq:dotE}), the pressure adds to the energy density,
because it also gravitates, and  thereby the energy flux is, in
reality, a flux of enthalpy. Owing to Lorentz invariance, the
enthalpy of the quantum vacuum, $\rho_{\Lambda} + p_{\Lambda}$,
vanishes identically; thus, the enthalpy of the vacuum in any
spatial three-volume must also vanish. Consequently, the
fluctuations of the energy flux of the vacuum are identically
zero.
\\   \

To apply the LL method of calculating the fluctuations of the
fluxes in a system the latter must be at thermodynamical
equilibrium or near to it. The second possibility means that the
system should evolve slowly. In our case, ``slowly" entails that
the rate by which the horizon area increases per unit of horizon
area  does not exceed its expansion rate, $3H$. A brief
calculation gives
\begin{equation}
\frac{\dot{\cal A}_{H}/{\cal A}_{H}}{3H} = \frac{\rho + p}{\rho}.
\label{eq:AdotA}
\end{equation}
The right-hand side of this equation is smaller than unity for a
$\Lambda$CDM universe and a universe dominated by quintessence and
pressureless matter, and equal to unity for the Einstein-de Sitter
universe. By contrast, it is $4/3$ for a radiation dominated
universe. Thus, we can safely apply the LL method to determine the
statistical average strength of the fluctuations of $ \, -
\dot{E}$ on the apparent horizon for various cases of interest.
\\  \

According to this method, if the flux $\dot{y_{i}}$ of a given
thermodynamic quantity, which evolves in a generic dissipative
process, is governed by $\, \dot{y}_{i} = \Sigma_{j} \Gamma_{ij}
Y_{j} \, + \, \delta\dot{y}_{i}\, $ and the entropy rate can be
written as $\, \dot{S} = \Sigma_{i} (\pm Y_{i} \dot{y}_{i})$,
where $\, Y_{i} = (\partial \dot{S}/\partial \dot {y}_{i})$, then
the second moments of the fluctuations of the fluxes obey $\,
<\delta \dot{y}_{i} \delta\dot{y}_{j}> = (\Gamma_{ij} \, + \,
\Gamma_{ji})\delta_{ij} \, \delta(t_{i} \, - \, t_{j})$, where the
angular brackets stand for statistical average with respect to the
reference state (namely, $\Sigma_{j} \Gamma_{ij} Y_{j}$), which is
taken to be steady or quasisteady and corresponds to the
systematic part of the flux. Obviously, $<\delta \dot{y}_{i}> =
0$.
\\  \

In the case at hand, we have just one flux, $\dot{y} = - \dot{E}$,
and the above equations imply
\begin{equation}
\dot{S}_{H} = \frac{1}{4 \ell^{2}_{P}} \, \dot{{\cal A}_{H}} = 2
\pi \, G \,\frac{{\cal A}^{2}_{H}}{\ell^{2}_{P}} \, H \, (\rho\, +
\, p)
\label{eq:dotSH}
\end{equation}
as well as
\begin{equation}
<(\delta(-\dot{E}))^{2}> = 2 \Gamma \, \delta(\tau) =  \frac{3 \,
\ell^{2}_{P}}{8 \, \pi^{2} G^{2}} \, H \, \frac{\rho \, + \,
p}{\rho} \, \delta(\tau)\; ,
\label{eq:deltadotE2}
\end{equation}
where $\tau$ is the time interval between two consecutive
measurements of $\dot E$ (say $t_i-t_j$ in the notation of Ref.
\cite{ll} recalled above).
\\  \

Notice that, due to the presence of the square of the Planck
length in the numerator of the latter equation, the average
strength of the fluctuations is minute, as expected. [Assuming the
Universe is well described at the background level  by the
$\Lambda$CDM model, currently $ <(\delta(-\dot{E}))^{2}>^{1/2}$
and $(-\dot{E})$ are, in natural units,  of the order of $\,
10^{-21}$ and $\, 10^{7}$, respectively]. In accordance with our
previous comment, they vanish, for an exact de Sitter universe, as
they should. This is quite reasonable, since the quantum vacuum is
continuous at the classical level, and therefore it does not
source classical fluctuations. Furthermore, because the right-hand
side of (\ref{eq:deltadotE2}) cannot be negative, and the Universe
is expanding, the null energy condition must be satisfied. This
directly excludes the dominance of phantom fields. This agrees
with the fact that these exotic fields are  disfavored by the
second law of thermodynamics \cite{grg-ninfa}, the occurrence of
quantum instabilities and other inherent maladies \cite{prd-cline,
ejp-sbisa, ejp-marius}. Likewise, a restriction arises on the
product $\, a^{2} \dot{H}$. The latter must fulfill $\, a^{2}
\dot{H} < k$. This is guaranteed when the spatial sections are
either spherical or flat; when they are hyperbolic, each
particular case should be studied on a single basis. Moreover, the
fact that the statistical averaged size of the fluctuations grows
with $\, H\, $ implies that the lower the scale factor, the lower
the area of the apparent horizon and the larger the strength of
the fluctuations. (For instance, in the $\Lambda$CDM universe, $\,
\rho = \rho_{\Lambda} + \rho_{m0} \, a^{-3}$, while $\, {\cal
A}_{H} \sim [\rho_{\Lambda} + \rho_{m0} \, a^{-3} + k \,
a^{-2}]^{-1}$). This result could have been anticipated on
physical grounds. It parallels the behavior of the fluctuations of
the fluxes in fluids (the smaller the volume of the fluid under
consideration, the stronger the fluctuations of the fluxes
\cite{ll}). Further, since the temperature of the horizon is
proportional to its surface gravity and this increases with the
Hubble factor \cite{rgcai}, so does the size of the fluctuations.
They behave, also in this regard, similarly to the statistical
fluctuations of normal systems not dominated by gravity.
\\  \

Clearly, the intensity of the random fluctuations of $\,
(-\dot{E}) \,$ should be fairly lower than $(- \dot{E})$ itself.
In other words,
\begin{equation}
\eta = \frac{ 3 \ell^{2}_{P}}{8 \pi^{2} G^{2}} \frac{H}{{\cal
A}^{2}_{H} \, \rho \, (\rho \, + \, p)} \, \delta(\tau) < 1.
\label{eq:eta}
\end{equation}
\\  \

At late times, the Universe must approach a state of maximum
entropy; this means that  it will get steadily dominated by the
cosmological constant, with $\, (\rho \, + \, p) \rightarrow 0$.
Accordingly, $\eta$ will grow. However, the reasonable condition
(\ref{eq:eta}) sets a generous upper bound on the fluctuations.
This is rather sensible because, as said above, in an exact de
Sitter expansion the energy flux vanishes identically and so do
its fluctuations.
\\  \

The aforesaid bound can be recast as a lower limit on the energy
flux,
\begin{equation}
(- \dot{E}) \; > \; \frac{3 \, \ell^{2}_{P}}{8 \pi^{2} \, G^{2}} \,
\frac{H}{{\cal A}_{H} \, \rho} \, \delta(\tau). \label{eq:fluxlimit}
\end{equation}
A stronger bound, not yet found, must hold in the quantum regime.
\\  \

Note that (\ref{eq:fluxlimit}) does not apply to the case of the
quantum vacuum itself. Indeed, as mentioned above, the latter
content \textemdash unlike  matter and radiation \textemdash is
neither discontinuous nor presents thermal motion.
\\  \

At early times (i.e., well before the vacuum energy started to
dominate) the energy density could (in principle) be so high that
the bound (\ref{eq:eta}) would be violated \textemdash recall that
$\, {\cal A}_{H} \propto \rho^{-1}$ and that $(\rho + p) \sim -
\dot{H} > 0$. However, if this were to occur, it would be very
likely  to happen  before the beginning of the matter era, whence
the near-equilibrium condition $\, (\rho + p)/\rho \leq 1$ would
not be met at that epoch, and the LL method would not apply,
according to Eq. (\ref{eq:AdotA}).
\\  \

The analysis carried out here can be readily extended to the case
in which there are sources of matter and/or radiation creation.
Then, the Universe should be treated  as an open system ``{\em \`a
la} Prigogine" \cite{grg-Ilya}. In this instance, the continuity
equation reads
\begin{equation}
\dot\rho \, + \, 3H\, (\rho+p)=\Gamma_{c} \,\rho\; ,
\label{continuity-gamma}
\end{equation}
where $\Gamma_{c}$ is the rate of creation of energy.
It obeys $\, 0 \leq \Gamma_{c}/3H < 1$ \cite{pla-calvao,
pla-winfried, prd-rafa,Harko:2015pma}. Then,
\begin{equation}
- \dot{E} = {\cal A}_{H} \, (\rho \, + \, p) \, H \, \tilde{r}_{H}
= \frac{{\cal A}_{H}}{3 H}\, \left[\Gamma_{c} \, \rho \, - \,
\frac{3}{4 \pi G} H \, (\dot{H} \, - \, k a^{-2}) \right] H \,
\tilde{r}_{H}.
\label{dotE1}
\end{equation}
Associated to this rate there is an extra, negative, pressure, $\,
p_{c} = - (\rho + p) \Gamma_{c}/3H$. However, as it is easy to
realize, when there is creation of particles, the energy flux is
augmented accordingly. However, as can be checked, the expression
for the statistical fluctuations of $(- \dot{E})$ formally
coincide with (\ref{eq:deltadotE2}), though $\, \rho$, $p$ and $\,
H$ will differ from the situation where $\Gamma_{c} = 0$.

\section{Conclusions}
We have evaluated the statistical average strength of the
classical fluctuations of energy flux on the apparent horizon of
an expanding FRW universe, namely, Eq. (\ref{eq:deltadotE2}). This
equation has a number of desirable features (e.g., the
fluctuations increase with the temperature of the horizon and
decrease with its area), thus showing clear similarities with the
typical fluctuations in systems in which gravity does not play a
main role, and they identically vanish when the expansion is
purely de Sitter. This indicates to us that the Universe, governed
by Einstein gravity, behaves as a normal thermodynamic system. It
should be interesting to study these fluctuations assuming the
Universe is described by any other reasonable theory of gravity
\cite{Tian:2014sca,delaCruzDombriz:2012xy}, which is left for a
subsequent work.

\section*{ Acknowledgments}
D.P. is grateful to the Departamento de F\'{i}sica e Instituto de
Astrof\'{\i}sica  e Ci\^{e}ncias do Espa\c{c}o do la Universidade
do Lisboa, where part of this work was done, for warm hospitality
and financial support. J.P.M. acknowledges the financial support
by Funda\c{c}\~ao para a Ci\^encia e a Tecnologia (FCT) through
research grant No. UID/FIS/04434/2013. This article is based upon
work from European Cooperation in Science and Technology (COST)
Action CA15117, CANTATA, supported by COST.

\end{document}